%% file: main.tex
\documentclass[lettersize,journal]{IEEEtran}
\usepackage{amsmath,amsfonts}
\usepackage{algorithmic}
\usepackage{algorithm}
\usepackage{array}
\usepackage[caption=false,font=normalsize,labelfont=sf,textfont=sf]{subfig}
\usepackage{stfloats}
\usepackage{url}
\usepackage{graphicx}
\usepackage{cite}
\usepackage[table]{xcolor}
\usepackage{tikz}
\usepackage{adjustbox}
\usepackage{multirow}
\usepackage{booktabs}
\usepackage[caption=false,font=normalsize,labelfont=sf,textfont=sf]{subfig}
\usepackage{caption}
\usetikzlibrary{
  arrows.meta,
  shapes,
  shadows,
  positioning,
  calc,
  backgrounds,
  fit,
  decorations.pathreplacing
}
\definecolor{accent}{HTML}{2B8CBE}
\definecolor{accent2}{HTML}{4DAF4A}
\definecolor{warm}{HTML}{F39C12}
\definecolor{muted}{HTML}{F2F2F2}
\definecolor{card}{HTML}{F7F9FB}
\definecolor{textcol}{HTML}{222222}

\begin{document}

\title{Lightweight Low-Frequency Non-Intrusive Load Monitoring via Karhunen-Loève Feature Distillation and Temporal Convolutional Networks}

\author{Sanka~Peeris$^{\dagger}$,
        Pandula~Thennakoon$^{\dagger}$,
        Praveen~Sumanasekara$^{\dagger}$,
        Roshan~Godaliyadda,~\IEEEmembership{Senior Member,~IEEE,}
        Parakrama~Ekanayake,~\IEEEmembership{Senior Member,~IEEE,}
        and~Janaka~Ekanayake,~\IEEEmembership{Fellow,~IEEE}
\thanks{This work has been submitted to the IEEE for possible publication. Copyright may be transferred without notice, after which this version may no longer be accessible.}
\thanks{$^{\dagger}$S. Peeris, P. Thennakoon and P. Sumanasekara contributed equally
to this work. (Corresponding author: Sanka Peeris.)}%
\thanks{S. Peeris is with the Multidisciplinary AI Research Centre (MARC),
University of Peradeniya, Peradeniya 20400, Sri Lanka
(e-mail: sankap@eng.pdn.ac.lk).}%
\thanks{P. Thennakoon is with Rice University, Houston, TX 77005 USA, and
also with Farbe Technologies, Sri Lanka (e-mail: ett6@rice.edu).}%
\thanks{P. Sumanasekara is with the University of Michigan, Ann Arbor,
MI 48109 USA, and also with the Multidisciplinary AI Research Centre (MARC),
University of Peradeniya, Peradeniya 20400, Sri Lanka
(e-mail: upraveen@umich.edu).}%
\thanks{R. Godaliyadda, P. Ekanayake, and J. Ekanayake are with the
Multidisciplinary AI Research Centre (MARC),
University of Peradeniya, Peradeniya 20400, Sri Lanka and the Department of Electrical
and Electronic Engineering, University of Peradeniya, Peradeniya 20400,
Sri Lanka (e-mail: roshangodd@ee.pdn.ac.lk; mpb.ekanayake@ee.pdn.ac.lk;
jbe@ee.pdn.ac.lk).}%
}
\maketitle

\begin{abstract}
Non-Intrusive Load Monitoring (NILM) is a critical enabler for demand-side management which allows for smart grid optimization. A deployment-ready NILM solution necessitates three key features: accurate disaggregation, low-frequency sampling, and low computational cost. This paper proposes a lightweight, edge-oriented NILM framework that combines activity-boundary-aware synthetic data augmentation with Karhunen--Lo\`{e}ve Transform (KLT) feature extraction and a Temporal Convolutional Network (TCN). The augmentation stage extracts and recombines complete or near-complete appliance activation segments so that the model can be trained without new fully labeled household aggregates. The KLT filter bank extracts decorrelated, appliance-informed temporal features which are then fed into a compact TCN along with the original aggregate for energy disaggregation. Evaluations on two standard NILM datasets, UK-DALE and REDD, at 10-s sampling and with leave-one-house-out testing, demonstrate that the proposed architecture achieves comparable or superior disaggregation accuracy relative to NILMFormer, BERT4NILM, BiLSTM, and BiGRU. The computational footprint of the framework is 31,009 parameters, the fewest of the five models compared, and on a single CPU thread it needs 2.78~ms per 1000 samples, 84--95\% less than the baselines. These results demonstrate a favorable accuracy-efficiency tradeoff for low-frequency smart-meter NILM.
\end{abstract}


\input{src/1_Introduction}
\input{src/2_Methodology}
\input{src/3_Results}
\input{src/4_Conclusion}

\bibliographystyle{IEEEtran}
\bibliography{references}

\end{document}

%% file: src/1_Introduction.tex
\section{Introduction}
\IEEEPARstart{W}{hile} rising energy demand and the need to incorporate renewables have placed growing importance on smart grid technology, gaining detailed information on electricity consumption remains a key challenge to its efficient operation. Advanced metering infrastructure increasingly provides low frequency aggregate power measurements at household endpoints, enabling non-intrusive load monitoring (NILM) for appliance-level disaggregation, contributing to demand-side management~\cite{CARRIEARMEL2013213}. However, a practically deployable NILM solution must combine accurate and reliable disaggregation, compatibility with low-frequency smart-meters, and reduced computational footprint. This work targets these constraints through an edge-oriented, low-frequency NILM framework that couples data-adaptive signal feature extraction with compact temporal modeling and synthetic data augmentation.

NILM is a widely accepted load monitoring technique that infers individual loads by considering the total aggregate power signal. This is in contrast to Intrusive Load Monitoring (ILM), where it is necessary to install power meters to each appliance individually. Since NILM only requires the total signal, it only requires one smart meter to be installed that is common to a set of appliances, which makes it more practical and economical compared to ILM. First proposed in 1992~\cite{192069}, the field has evolved, yielding a range of solutions that can determine the operating status and energy consumption of individual electrical loads by analyzing total current, voltage, or power signals. In addition to its role in optimizing smart grids, NILM also finds application in power consumption surveys, home automation, as well as the formulation of energy policies and regulations.

Existing literature focuses on different aspects of this, such as state estimation~\cite{CALI:A-Comprehensive-Multi-Type, fang_multi-length_2025}, event detection~\cite{8745491, 9634017}, and estimating the power consumption of individual appliances~\cite{liang_heterogeneous_2025, dash_multi-task_2024, welikala_incorporating_2019, Zijun-Su:Zero-Shot-Nonintrusive-Load-Monitoring}. A majority of the recent algorithms focus on supervised learning techniques, where data containing both the aggregate signal as well as the disaggregated individual appliance signals are used to train these algorithms. However, a subset of the research involves unsupervised or semisupervised algorithms to alleviate the need for labeled data. A real-world deployable NILM method needs to satisfy all three of the below conditions.

\begin{enumerate}
    \item Robust accuracy and high disaggregation performance to ensure reliability.
    \item Must be compatible with existing smart meter infrastructure that runs on low-frequency sampling.
    \item Minimal memory footprints and low computational complexity for edge deployment.
\end{enumerate}

Current NILM methods mainly split into two categories based on their design choice, namely signal processing methods and deep learning based methods. Classical signal processing methods~\cite{10044234, WU2021111025, 7317784} are fast and light weight while being reasonably accurate, making them suitable for edge deployment. On the other hand, deep learning methods surpass traditional signal processing methods on accuracy and disaggregation performance. However, recent attention/sequence models can increase parameter, memory, and inference costs.

In addition to this, one can identify two mainstream approaches based on the sampling frequency used for the signals, namely high-frequency (kHz) sampling and low-frequency (seconds to minutes) sampling. Some modern algorithms rely on high-frequency data with sampling rates in the kHz range \cite{DeepDFML,ST-NILM,NILM-Multivariate-DNN-HF}, which allows for the exploitation of harmonics as well as Voltage-Current waveforms for the NILM problem. Despite the fine-grained information available, the practical limitations of existing smart meters, as well as the computing power required to process and transmit these signals at these high frequencies, remain a challenge. Because of this, NILM algorithms that rely on low-frequency signals sampled at the seconds to minutes range are an active area of research, as they are better suited for the deployment conditions of many existing smart meter systems. This work focuses on the problem of appliance-level disaggregation using low-frequency aggregate power signals.

\IEEEpubidadjcol 
Deep learning based data driven NILM models are heavily constrained by the lack of diversity in publicly available datasets. Because models are trained on a limited number of households, and therefore on a narrow set of appliance makes and usage habits, they often transfer poorly to unseen households~\cite{wu_harnessing_2025, pusceddu_fine-tuning_2025, liu2024nonintrusiveloadmonitoringsmart}. Moreover, gathering appliance-resolved aggregate measurements is a slow and expensive process that cannot keep pace with an evolving appliance ecosystem. Because of this, trained models can degrade noticeably on households they have not seen during training~\cite{10.1145/3580305.3599910}. We therefore explore synthetic data generation as one way to address this, using a pipeline that turns easily obtainable, short single-appliance recordings into large and diverse aggregate training data. Rather than generating abstract waveforms, our method uses \emph{activity-boundary-aware sampling}: segment start and end points are chosen only within low-power, inactive regions of appliance traces, so that each extracted segment contains a complete or near-complete activation rather than an arbitrary cut through an appliance's operation. By concatenating these genuine activation events with controlled idle intervals and summing across devices, we generate synthetic aggregate streams that retain the measured activation shapes of each appliance.

Current state-of-the-art deep learning models~\cite{liang_heterogeneous_2025, yue_bert4nilm_2020, huzzat_gru-bert_2025, RONG2025115334, electronics15020453, 10887006, dash_multi-task_2024, Zijun-Su:Zero-Shot-Nonintrusive-Load-Monitoring} have increasingly relied on larger architectures, specifically Transformers and hybrid networks combining Convolutional Neural Networks (CNNs), Recurrent Neural Networks (RNNs), and self-attention mechanisms. While these models achieve state-of-the-art accuracy by successfully capturing complex, long-range temporal dependencies in power signals, they can require substantially more parameters, memory and computation than simpler models. This matters for deployment on edge devices, which have limited memory, power and processing capacity~\cite{li_lightweight_2025}. Relying on cloud computing to process these complex models introduces latency, privacy concerns, and high bandwidth costs. This motivates lightweight models that retain competitive accuracy.

Recent work has addressed these requirements along three broad approaches: making models small enough to run on end devices~\cite{li_lightweight_2025}, designing architectures and features suited to low-frequency or scarce data~\cite{liang_heterogeneous_2025, yu_multi-time-scale_2024, fang_multi-length_2025}, and improving generalization through synthetic training data, label-free transfer or added usage context~\cite{CALI:A-Comprehensive-Multi-Type, Zijun-Su:Zero-Shot-Nonintrusive-Load-Monitoring, chen_scene-aware_2026}.

In this work, we combine training data assembled from measured activations with boundary-aware cropping, a Karhunen-Lo\`eve Transform (KLT) feature front-end and a compact Temporal Convolutional Network (TCN), for continuous power disaggregation at 10-s resolution on households unseen during training. KLT was first applied to NILM by Dinesh et al.~\cite{dinesh_residential_2015, dinesh_residential_2016}. One of the greatest challenges in practical NILM is that unique appliance signatures are often obscured when data is collected at low sampling rates by standard smart meters. The KLT addresses this by representing windows of the power signal in a data-learned basis, where a few uncorrelated components capture most of the variance. Building on these findings, we use the KLT as the feature-extraction stage of our pipeline, which lets a compact network reach competitive accuracy.

In summary, the main contributions of this work are as follows.

\begin{enumerate}
    \item We propose a low-frequency lightweight NILM framework that achieves a strong accuracy-efficiency tradeoff: 31,009 parameters, 2.78 ms per 1,000 samples on one CPU thread. The framework has competitive performance across multiple appliance/metric cases, without dependence on large bidirectional or attention-heavy architectures.

    \item A signal-processing based feature extraction pipeline where raw power signals are fed to a KLT filter bank to give the model added appliance-specific pattern visibility.

    \item A boundary-aware synthetic data generator that extracts complete or near-complete appliance activations and recombines them into labelled aggregate training sequences.

    \item A leave-one-house-out evaluation of the proposed architecture on UK-DALE and REDD datasets, showing competitive or superior performance and the lowest measured CPU latency among the evaluated baselines. Both benchmarks use 10-second sampling for realistic low-frequency operation in the field.
\end{enumerate}

%% file: src/2_Methodology.tex
\section{Methods}

\subsection{Sampling Algorithms for Synthetic Data Generation}

\subsubsection{Synthetic Data Augmentation for Incremental Appliance Integration}

\begin{figure*}[!t]
\centering
\includegraphics[width=\textwidth]{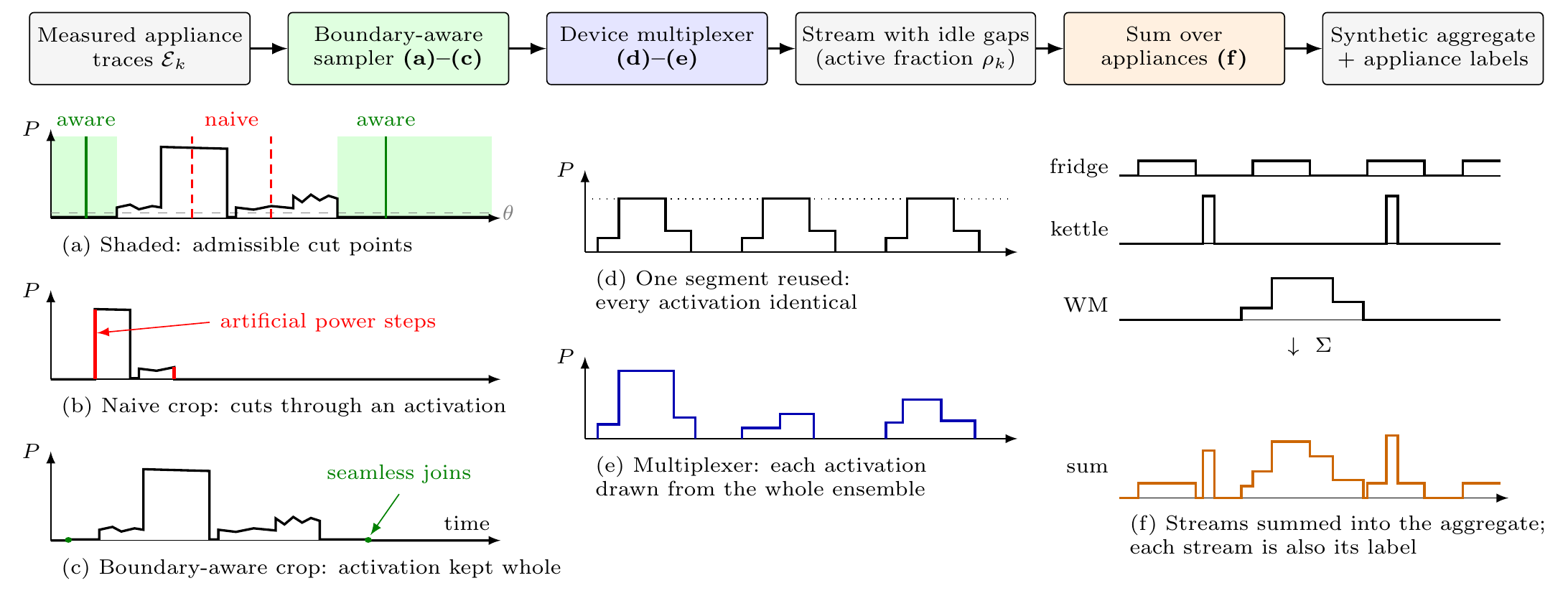}
\caption{Synthetic training-data pipeline. (a)--(c) The boundary-aware sampler cuts only in inactive regions, so activations stay whole. (d)--(e) The multiplexer draws each activation from the whole measured ensemble. (f) The per-appliance streams are summed into the aggregate and also serve as labels. Traces are illustrative.}
\label{fig:sampler_contrast}
\end{figure*}

A practical limitation of NILM deployment is that labeled household-level aggregate datasets do not grow at the same rate as the appliance ecosystem. New appliance variants are introduced continuously, while collecting aggregate measurements with appliance-resolved labels remains expensive and slow. In contrast, short device-level recordings are comparatively easy to obtain. This motivates a synthetic data generation stage that transforms a small set of single-appliance recordings into a large set of aggregate training examples. The full pipeline is shown in Fig.~\ref{fig:sampler_contrast}

Let the selected appliance set be
\begin{equation}
\mathcal{A} = \{1,\dots,N_{\mathrm{app}}\}.
\end{equation}
For each appliance $k \in \mathcal{A}$, assume a manufacturer or curator provides a small ensemble of measured real-power traces (in our experiments, activations from the training-house submeters)
\begin{equation}
\mathcal{E}_k = \left\{ s_k^{(1)}, s_k^{(2)}, \dots, s_k^{(M_k)} \right\},
\end{equation}
where each $s_k^{(m)} \in \mathbb{R}^{T_m}$ is a single-appliance recording. The objective is to generate synthetic appliance streams $x_k(t)$ and their aggregate
\begin{equation}
x_{\mathrm{agg}}(t) = \sum_{k=1}^{N_{\mathrm{app}}} x_k(t), \qquad t=1,\dots,T,
\end{equation}
for arbitrary sequence length $T$, while preserving realistic appliance behavior.

Unlike parametric waveform simulators, the proposed sampler does not invent appliance signatures from abstract models. Instead, it reuses measured activation patterns. For each trace $s_k^{(m)}$, low-power regions are identified as candidate inactive boundaries. Denote these boundary indices by $\Omega_k^{(m)}$. A candidate active segment is generated by sampling $a,b \in \Omega_k^{(m)}$ with $a<b$ and extracting
\begin{equation}
r_k(t) = s_k^{(m)}(t), \qquad t \in [a,b],
\end{equation}
subject to a minimum non-trivial activity criterion over $[a,b]$. This boundary-aware extraction favors complete or near-complete activation events and reduces arbitrary mid-operation cuts (Fig.~\ref{fig:sampler_contrast}a--c).

For appliance $k$, a long synthetic stream of length $T$ is then formed by concatenating sampled active segments and zero-valued idle intervals. Let $\rho_k \in [0,1]$ denote the desired active fraction for appliance $k$. The target active duration is
\begin{equation}
L_k = \lfloor \rho_k T \rfloor,
\end{equation}
and active segments are sampled until their total duration reaches approximately $L_k$. The remaining duration
\begin{equation}
Z_k = T - L_k
\end{equation}
is allocated to idle intervals distributed between active segments, yielding
\begin{equation}
x_k = [\,0\cdots0,\ r_{k,1},\ 0\cdots0,\ r_{k,2},\ \dots,\ r_{k,n_k},\ 0\cdots0\,].
\end{equation}
Each appliance stream is generated independently, and aggregate samples are obtained by summation across appliances. Each activation is drawn from the whole ensemble rather than from a single segment (Fig.~\ref{fig:sampler_contrast}d--e), and the per-appliance streams are summed into the aggregate, with each stream serving as its appliance's label (Fig.~\ref{fig:sampler_contrast}f). In our experiments, synthetic windows replace the real training windows one for one, so no real aggregate is seen during training (Section~\ref{subsec:ablation_study})

This construction is especially useful in low-data NILM settings for four reasons. First, it preserves device realism because active segments are copied from measured traces, retaining waveform shape, transients, and intra-cycle variability that are often important for disaggregation. Second, it avoids destructive segmentation: boundary selection in low-power regions reduces physically implausible fragments caused by random cropping. Third, it adds sequence-level variety without altering the measured waveforms: even with small $M_k$, recombining segments with different idle intervals and overlap patterns yields aggregates that differ from the original recordings. Fourth, it provides explicit prevalence control through $\rho_k$, enabling either realistic duty-cycle modeling or targeted balancing for underrepresented appliances.

\begin{figure*}[t]
    \centering
    \begin{adjustbox}{max width=\textwidth}
    \begin{tikzpicture}[
        font=\footnotesize,
        >=latex,
        line width=0.6pt,
        node distance=4mm,
        proc/.style={
            draw,
            rectangle,
            text width=2.55cm,
            minimum height=2.2cm,
            inner sep=2pt,
            align=center,
            fill=black!3,
            execute at begin node={\hyphenpenalty=10000}
        },
        emph/.style={
            proc,
            fill=black!10,
            very thick
        },
        note/.style={
            draw,
            rectangle,
            rounded corners=1.5pt,
            text width=2.55cm,
            inner sep=2pt,
            align=center,
            fill=black!2,
            font=\scriptsize,
            execute at begin node={\hyphenpenalty=10000}
        }
    ]


    \node[emph] (submit)
    {\textbf{Manufacturer\\Trace Submission}\\
    \scriptsize Small single-appliance traces of a new device/variant};

    \node[proc, right=of submit] (repo)
    {\textbf{Appliance Trace Repository}\\
    \scriptsize Existing library + appended submissions};

    \node[emph, right=of repo] (synth)
    {\textbf{Synthetic\\Augmentation Engine}\\
    \scriptsize Aggregates + labels, controllable activity ($\rho_k$)};

    \node[proc, right=of synth] (data)
    {\textbf{Training Dataset Update}\\
    \scriptsize Existing corpus + synthetic data};

    \node[proc, right=of data] (update)
    {\textbf{NILM Model Update}\\
    \scriptsize Retrain / fine-tune on new devices};

    \node[proc, right=of update] (deploy)
    {\textbf{Deployment}\\
    \scriptsize Release updated model to field meters/platform};


    \draw[->] (submit) -- (repo);
    \draw[->] (repo)   -- (synth);
    \draw[->] (synth)  -- (data);
    \draw[->] (data)   -- (update);
    \draw[->] (update) -- (deploy);


    \node[note, above=4mm of data] (n4)
    {No new fully labeled household dataset required};
    \draw (n4) -- (data);

    \coordinate (loop) at ([yshift=4mm]n4.north);
    \draw[->,semithick,rounded corners=6pt]
        (deploy.north) -- (deploy.north |- loop)
        -- node[above, font=\scriptsize]
           {Future submissions trigger the next update cycle}
           (submit.north |- loop)
        -- (submit.north);


    \node[note, below=4mm of submit] (n1) {Low burden for manufacturers};
    \draw (n1) -- (submit);

    \node[note, below=4mm of synth] (n2)
    {Bridge: limited traces $\rightarrow$ large supervised NILM data};
    \draw (n2) -- (synth);

    \node[note, below=4mm of update] (n3) {Pathway to new appliance variants};
    \draw (n3) -- (update);

    \end{tikzpicture}
    \end{adjustbox}

    \caption{Automated manufacturer-assisted synthetic data augmentation pipeline for continual NILM model updates. Manufacturer-provided appliance traces are incrementally integrated into a repository, transformed into large synthetic aggregate-label pairs, and used to update and redeploy the NILM model without full household-level relabeling.}

    \label{fig:proposed_model_update_pipeline}
\end{figure*}
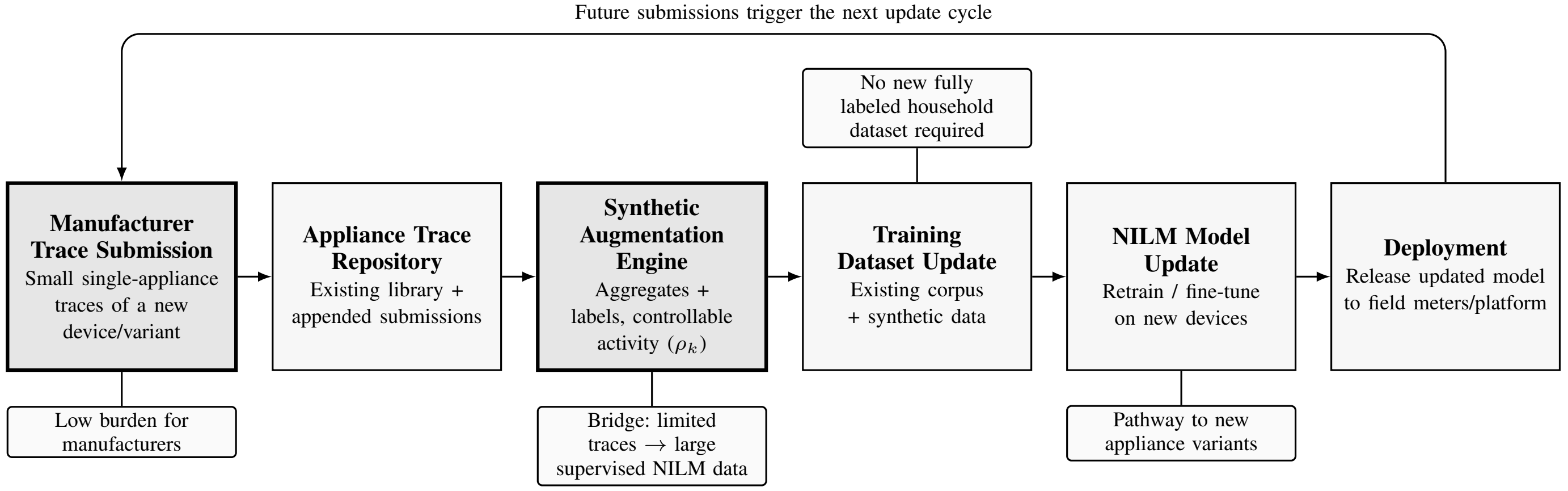

Within the proposed pipeline, this mechanism provides a pathway to post-deployment updates. When a manufacturer submits a small set of device-level traces for a new product or revision, those traces are appended to the corresponding ensemble $\mathcal{E}_k$. New synthetic aggregates that include the submitted device can then be generated automatically, and the NILM model can be retrained or fine-tuned in the next release cycle. Consequently, support for emerging appliances can reduce dependence on collecting large new household-level labeled datasets (See Fig.~\ref{fig:proposed_model_update_pipeline}).

In summary, the proposed synthetic augmentation strategy combines realism, scalability, and adaptability: it is realistic because it is anchored in measured behavior, scalable because it expands small device-level ensembles into large aggregate corpora, and adaptable because newly submitted appliance traces can be integrated into continuous model update workflows.

\subsection{Karhunen-Loève Transform (KLT) based Feature Engineering and Model}

\begin{figure*}[!t]
    \centering
    \includegraphics[width=1.0\linewidth]{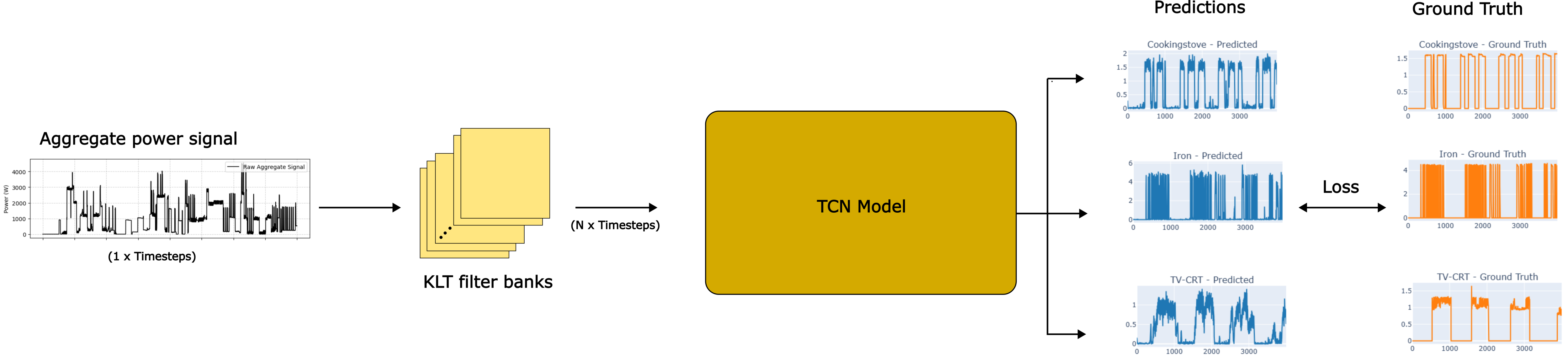}
    \caption{Model Architecture. Output traces are illustrative.}
    \label{fig:Model}
\end{figure*}

The proposed feature extraction pipeline utilizes an empirically derived KLT to map the aggregate power signal into a discriminative, multi-channel feature space. Unlike generic basis transforms, the KLT eigenfilters are computed directly from an ensemble of isolated appliance activation signatures from the training houses, so that the basis adapts to the specific transient and steady-state characteristics of the target load profiles. This process and the proposed model is depicted in Fig.~\ref{fig:Model}

Before feature extraction, the discrete-time aggregate power signal, $x[n]$, is normalized using a global constant $C$, the maximum aggregate power observed in the training houses ($C = 6852$~W for UK-DALE and $6822.5$~W for REDD), to stabilize the gradient descent during network training:
\begin{equation}
    x_{norm}[n] = \frac{x[n]}{C}
\end{equation}

The ground truth power sequence of the target appliance, $y_k[n]$, is scaled by the same constant, so that predictions and inputs share one power scale:
\begin{equation}
    y_{k,norm}[n] = \frac{y_k[n]}{C}
\end{equation}

To capture the underlying physical characteristics of the devices, the KLT basis is not derived from the continuous aggregate signal, but rather from a structured matrix of isolated appliance signatures. 

Let $\mathcal{S}$ denote the pooled set of isolated activation segments of all target appliances, extracted from the training houses. Each segment is standardized to zero mean and unit variance, and every length-$W$ sliding window of a standardized segment forms a column vector $\mathbf{s}_j \in \mathbb{R}^W$, $j = 1, \ldots, J$. The $W \times W$ empirical covariance matrix $\mathbf{C}_s$ is computed over this ensemble of windows:
\begin{equation}
    \mathbf{C}_s = \frac{1}{J} \sum_{j=1}^{J} \mathbf{s}_j \mathbf{s}_j^T
\end{equation}
with $W = 10$, so that a single basis is shared by all appliances.

Eigendecomposition is performed on $\mathbf{C}_s$ to yield an orthonormal basis:
\begin{equation}
    \mathbf{C}_s \mathbf{v}_p = \lambda_p \mathbf{v}_p, \quad p = 1, 2, \ldots, W
\end{equation}
The eigenvectors $\mathbf{v}_p$ are sorted by their corresponding eigenvalues $\lambda_p$ in descending order. To achieve optimal energy compaction while controlling the computational footprint of the network, we truncate the expansion order to $N$ (where $N=10$). The top $N$ eigenvectors are then deployed as the impulse responses of a Finite Impulse Response (FIR) filter bank:
\begin{equation}
    h_p[\tau] = \mathbf{v}_p[\tau], \quad \tau \in \{0, 1, \ldots, W-1\}, \quad p \in \{1, \ldots, N\}
\end{equation}

The normalized aggregate signal $x_{norm}[n]$ is projected onto the $N$ eigen filters over a sliding window to generate an $N$-dimensional feature space:
\begin{equation}
    f_p[n] = \sum_{\tau=0}^{W-1} h_p[\tau]\, x_{norm}[n+\tau]
\end{equation}
where $x_{norm}$ is zero-padded at the end of the sequence.

Crucially, to preserve the low-frequency absolute amplitude information that may be attenuated by the KLT projection, the original normalized aggregate signal is explicitly concatenated to the filtered channels. This yields the final $(N+1)$-channel multivariate input tensor $\mathbf{X}_{in}[n]$ passed to the TCN:
\begin{equation}
    \mathbf{X}_{in}[n] =
    \begin{bmatrix} f_1[n] \\ \vdots \\ f_N[n] \\ x_{norm}[n] \end{bmatrix}
    \in \mathbb{R}^{N+1}
\end{equation}

\begin{figure}[!t]
    \centering
    \includegraphics[width=1.0\linewidth]{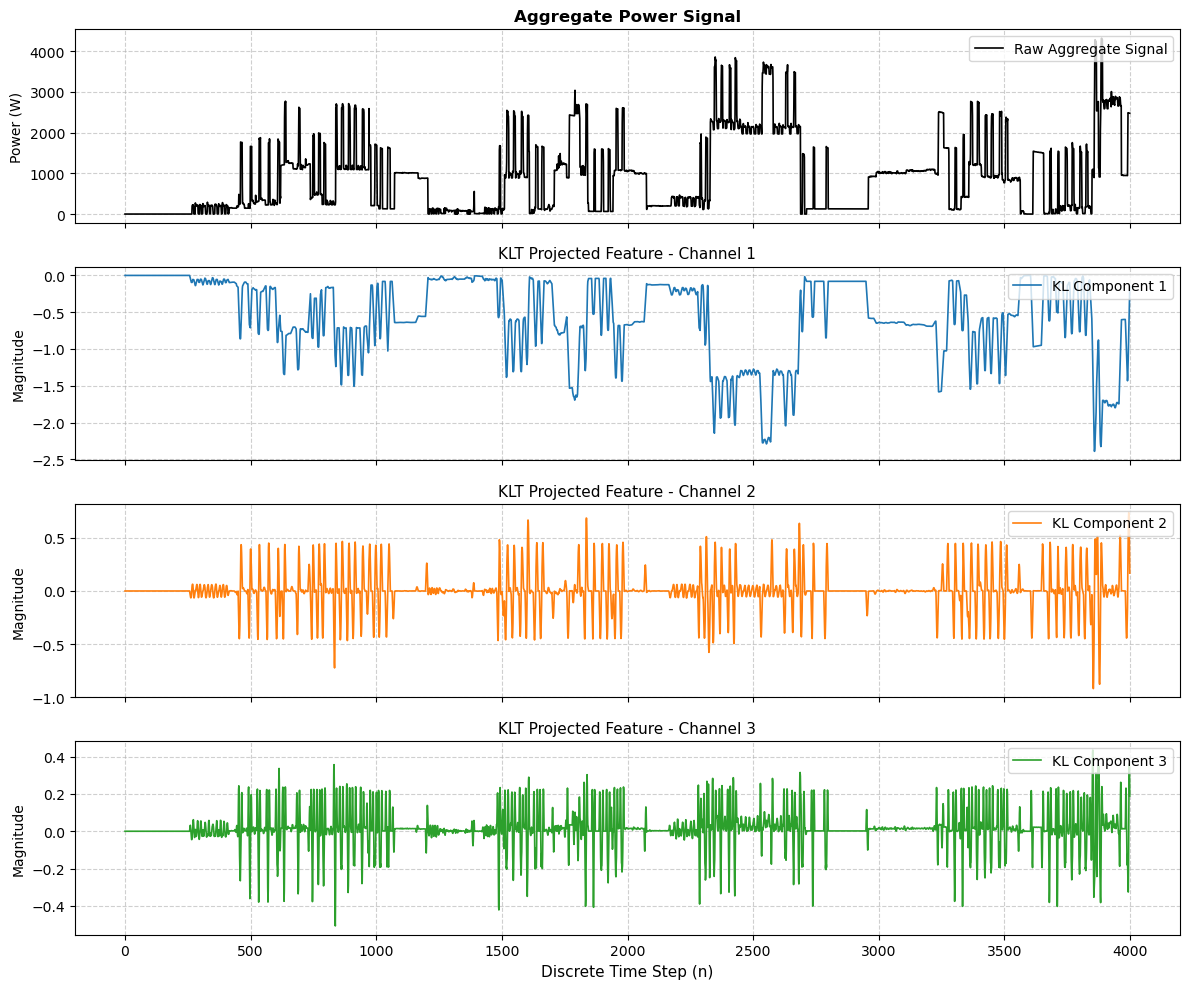}
    \caption{Example KLT-projected channels from an aggregate power segment.}
    \label{fig:KLT}
\end{figure}

Figure ~\ref{fig:KLT} shows a part of this input tensor to the model.

\subsubsection{Inference Constraints}
During the evaluation phase, strict parameter isolation is enforced. The global scaling constant and the $N$ eigen filters $h_p$, both derived from the training houses, are kept frozen. Unseen aggregate sequences are subjected to the exact same transformation and concatenation pipeline before being processed by the trained TCN.

\subsection{Temporal Convolutional Network Architecture}

To map the projected $(N+1)$-dimensional multivariate feature space into appliance-specific energy estimates, we employ a sequence-to-sequence TCN. The architecture is specifically designed to capture the short-term and multi-scale temporal structure of load profiles without the vanishing gradient limitations inherent to recurrent neural networks.

The TCN backbone comprises a stack of four residual blocks. Each block utilizes dilated 1D convolutions with a kernel size of $4$ and a consistent hidden dimension of $32$ filters.

For a given residual block $l$, the operation on the hidden state $\mathbf{H}^{(l)}$ with dilation factor $d$ is formulated as:
\begin{equation}
    \mathbf{H}^{(l+1)}[n] = \text{ReLU}\left( \mathbf{H}^{(l)}[n] + \mathcal{F}\left(\mathbf{H}^{(l)}[n], \boldsymbol{\Phi}^{(l)}, d\right) \right)
\end{equation}
where $\mathcal{F}$ represents the sequence of dilated convolutions, weight normalization, and dropout. The final layer of the network applies a linear $1 \times 1$ convolutional projection to reduce the $32$-dimensional hidden representation to a single output channel, since a separate model is trained for each target appliance. No non-linear activation is applied at this output stage, yielding unbounded regression outputs $\hat{\mathbf{y}}_{norm} \in \mathbb{R}^{T}$.

\subsubsection{Objective Function}

The network is trained to minimize the mean squared error between the predicted and ground-truth power sequences over a batch of $B$ sequences of length $T$:
\begin{equation}
    \mathcal{L}_{MSE} = \frac{1}{B \cdot T} \sum_{b=1}^{B} \sum_{n=1}^{T} \left( \hat{y}_{norm}^{(b)}[n] - y_{norm}^{(b)}[n] \right)^2
\end{equation}
The same loss is used for every baseline, so that differences between models reflect architecture and input features rather than training objective.

\subsubsection{Model Inference and Post-Processing}

During the operational inference phase, the model maps a continuous, unseen aggregated power sequence $x_{inf}[n]$ into individual physical power estimates.

First, the sequence is subjected to the frozen deterministic preprocessing pipeline: it is scaled by the global constant $C$ and passed through the frozen empirical Karhunen-Loève filter bank to yield the input tensor $\mathbf{X}_{inf}[n] \in \mathbb{R}^{N+1}$. 

The forward pass through the trained TCN generates the normalized prediction $\hat{y}_{norm}[n]$, which is rescaled to watts by the global constant $C$. Predictions below the appliance's activation threshold $\theta_k$ (Table~\ref{tab:thresholds}) are then set to zero, which also removes any negative outputs:
\begin{equation}
    \hat{y}_k[n] =
    \begin{cases}
        C\,\hat{y}_{norm}[n], & C\,\hat{y}_{norm}[n] \ge \theta_k \\
        0, & \text{otherwise}
    \end{cases}
\end{equation}

%% file: src/3_Results.tex
\section{Experiments}

\subsection{Experimental Setup}

\begin{table}[!t]
\centering
\caption{Hardware, Software and Training Configuration}
\label{tab:experimental_setup}
\begin{tabular}{@{}ll@{}}
\toprule
\textbf{Parameter} & \textbf{Value / Configuration} \\ \midrule
\multicolumn{2}{@{}l}{\textbf{Hardware and Software}} \\
CPU & Intel Xeon w7-3565X (32 cores) \\
GPU & NVIDIA RTX 6000 Ada Generation (48 GB) \\
RAM & 503 GiB \\
OS / Python & Ubuntu 24.04, Python 3.12 \\
Framework & PyTorch 2.12.0 (CUDA 13.0) \\ \midrule
\multicolumn{2}{@{}l}{\textbf{Model}} \\
KLT filter bank & $W=10$ taps, $N=10$ eigen-filters \\
Input channels & 11 (10 KLT + raw aggregate) \\
TCN & 4 residual blocks $\times$ 32 channels \\
Kernel size / dilations & 4 / 1, 2, 4, 8 \\
Normalization / dropout & Weight normalization / 0.1 \\
Output & 1 channel (one model per appliance) \\ \midrule
\multicolumn{2}{@{}l}{\textbf{Training}} \\
Loss & Mean squared error (MSE) \\
Optimizer & AdamW (lr $1\times10^{-4}$, weight decay 0) \\
LR scheduler & ReduceLROnPlateau (patience 3, factor 0.1) \\
Batch size & 64 \\
Epochs (UK-DALE) & 3, fixed \\
Epochs (REDD) & $\le$100, early stopping (patience 8) \\Gradient clipping & 1.0 (max norm) \\
Window length & 128, 256, 512 samples (non-overlapping) \\
Seeds & 0, 1, 2 \\ \bottomrule
\end{tabular}
\end{table}

\begin{table}[!t]
\centering
\caption{Dataset Partitioning and Sampling Configuration}
\label{tab:dataset_splits}
\begin{tabular}{@{}lllc@{}}
\toprule
\textbf{Dataset} & \textbf{Training Houses} & \textbf{Testing Houses} & \textbf{Sampling Rate} \\ \midrule
\textbf{UK-DALE} & 1, 3, 4, 5 & 2 & 10s \\
\textbf{REDD}    & 2, 3, 4, 5, 6 & 1 & 10s \\
\bottomrule
\end{tabular}
\end{table}

\begin{table}[!t]
\centering
\caption{Appliance Activation Thresholds}
\label{tab:thresholds}
\begin{tabular}{@{}lc@{}}
\toprule
\textbf{Appliance} & \textbf{Activation Threshold (W)} \\ \midrule
Kettle & 2000 \\
Microwave & 200 \\
Refrigerator & 50 \\
Washing Machine & 20 \\
Dishwasher & 10 \\ \bottomrule
\end{tabular}
\end{table}

\begin{table*}[!t]
\centering
\caption{Performance Comparison on UK-DALE and REDD. BA and F1 in \%, MAE in W. Best per appliance and metric in bold; the proposed method is shaded.}
\label{table:performance_comparison}
\begin{tabular}{c|c|ccccc|cccc}
\hline
\multirow{2}{*}{Metrics} & \multirow{2}{*}{Methods} & \multicolumn{5}{c|}{UK-DALE} & \multicolumn{4}{c}{REDD} \\ \cline{3-11}
 & & WM & DW & Kettle & Micro. & Fridge & WM & DW & Micro. & Fridge \\ \hline
\rowcolor{gray!15} \multirow{5}{*}{BA} & Proposed & 88.50\% & 68.58\% & 82.54\% & 95.63\% & \textbf{95.69\%} & 85.49\% & 83.06\% & \textbf{92.51\%} & 94.20\% \\

 & NILMFormer\cite{petralia_nilmformer_2025} & \textbf{90.93\%} & 86.37\% & \textbf{96.87\%} & 79.10\% & 95.06\% & \textbf{87.30\%} & 79.20\% & 85.39\% & 91.27\% \\
 & BERT4NILM\cite{yue_bert4nilm_2020} & 77.12\% & \textbf{86.72\%} & 92.82\% & \textbf{97.93\%} & 95.62\% & 86.36\% & \textbf{88.60\%} & 84.23\% & \textbf{95.42\%} \\
 & BiLSTM\cite{10.1145/2821650.2821672} & 80.39\% & 76.88\% & 74.18\% & 57.11\% & 65.74\% & 75.57\% & 74.96\% & 60.04\% & 79.09\% \\
& BiGRU\cite{Precioso2023} & 83.39\% & 72.82\% & 85.14\% & 52.39\% & 50.84\% & 86.16\% & 83.48\% & 50.00\% & 52.27\% \\
 \hline
\rowcolor{gray!15} \multirow{5}{*}{F1} & Proposed & 23.12\% & 32.59\% & 77.27\% & \textbf{73.94\%} & 94.87\% & 34.23\% & 42.79\% & 47.78\% & 87.42\% \\
 & NILMFormer\cite{petralia_nilmformer_2025} & \textbf{28.41\%} & 46.06\% & \textbf{95.80\%} & 49.18\% & 94.24\% & 22.37\% & 26.99\% & \textbf{60.50\%} & 80.54\% \\
 & BERT4NILM\cite{yue_bert4nilm_2020} & 18.00\% & \textbf{56.34\%} & 90.74\% & 67.42\% & \textbf{95.00\%} & \textbf{44.12\%} & \textbf{61.40\%} & 49.79\% & \textbf{88.88\%} \\
 & BiLSTM\cite{10.1145/2821650.2821672} & 7.20\% & 18.51\% & 55.51\% & 10.33\% & 38.37\% & 12.57\% & 23.23\% & 13.52\% & 57.01\% \\
& BiGRU\cite{Precioso2023} & 8.08\% & 27.74\% & 78.10\% & 6.16\% & 7.71\% & 20.06\% & 24.68\% & 0.00\% & 15.33\% \\
 \hline
\rowcolor{gray!15} \multirow{5}{*}{MAE} & Proposed & 35.16 & 40.63 & 8.49 & \textbf{2.84} & 12.61 & \textbf{20.63} & 20.65 & 18.46 & 23.95 \\
 & NILMFormer\cite{petralia_nilmformer_2025} & \textbf{9.13} & \textbf{16.22} & \textbf{3.00} & 5.88 & 11.30 & 33.97 & 22.84 & \textbf{13.15} & 28.59 \\
 & BERT4NILM\cite{yue_bert4nilm_2020} & 41.64 & 34.92 & 5.98 & 3.83 & \textbf{11.15} & 25.20 & \textbf{13.40} & 14.02 & \textbf{16.80} \\
 & BiLSTM\cite{10.1145/2821650.2821672} & 35.72 & 44.48 & 10.81 & 6.05 & 29.91 & 53.50 & 35.39 & 17.62 & 33.49 \\
 & BiGRU\cite{Precioso2023} & 31.37 & 35.97 & 7.47 & 6.15 & 38.36 & 47.02 & 27.21 & 15.91 & 52.15 \\
 \hline
\end{tabular}
\end{table*}

\begin{figure*}[!t]
\centering
\includegraphics[width=0.78\textwidth]{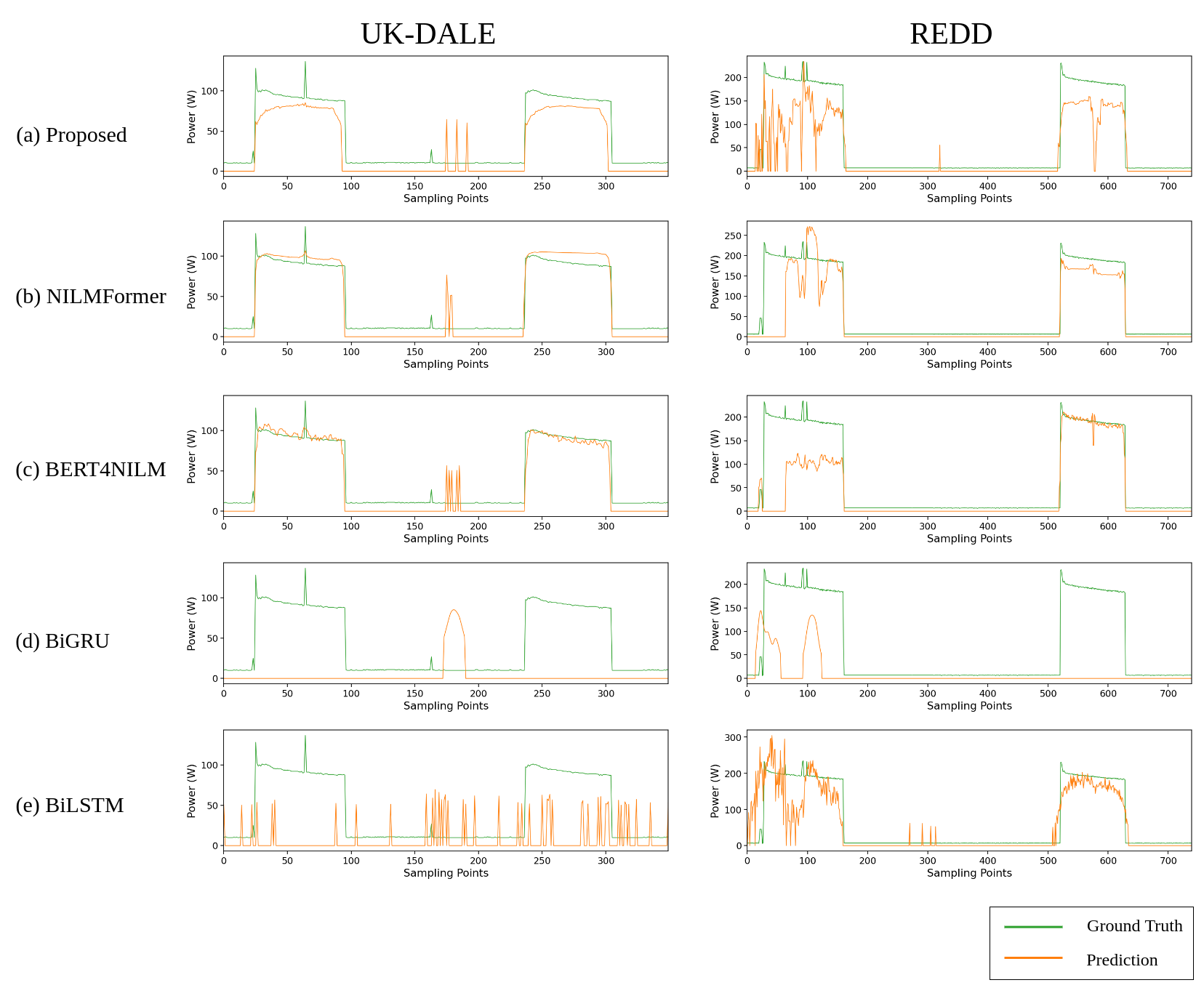}
\caption{Time-series comparison of ground truth vs. predicted power consumption for Fridge in the UK-DALE and REDD datasets}
\label{fig:sota-comparison}
\end{figure*}

\begin{table*}[!t]
\centering
\caption{Ablation of the KLT front end (KL) and the boundary-aware synthetic augmentation (Aug). TCN + KL + Aug is the proposed model.}
\label{table:ablation_study}
\begin{tabular}{c|c|ccccc|cccc}
\hline
\multirow{2}{*}{Metrics} & \multirow{2}{*}{Methods} & \multicolumn{5}{c|}{UK-DALE} & \multicolumn{4}{c}{REDD} \\ \cline{3-11}
 & & WM & DW & Kettle & Micro. & Fridge & WM & DW & Micro. & Fridge \\ \hline
\multirow{4}{*}{BA}
 & TCN & 78.07\% & 68.94\% & 55.64\% & 65.79\% & 66.22\% & \textbf{86.90\%} & 57.28\% & 50.46\% & 68.46\% \\
 & TCN + KL & 84.83\% & \textbf{73.39\%} & 81.63\% & 93.42\% & 95.13\% & 84.43\% & 80.91\% & 91.66\% & \textbf{94.81\%} \\
 & TCN + Aug & 76.54\% & 63.43\% & 51.90\% & 81.48\% & 65.76\% & 85.73\% & 57.32\% & 55.04\% & 63.82\% \\
 & TCN + KL + Aug & \textbf{88.50\%} & 68.58\% & \textbf{82.54\%} & \textbf{95.63\%} & \textbf{95.69\%} & 85.49\% & \textbf{83.06\%} & \textbf{92.51\%} & 94.20\% \\
 \hline
\multirow{4}{*}{F1}
 & TCN & 18.06\% & 23.98\% & 17.36\% & 20.18\% & 40.96\% & 20.27\% & 16.52\% & 1.50\% & 49.21\% \\
 & TCN + KL & 19.19\% & \textbf{34.90\%} & 75.46\% & 56.58\% & 94.21\% & 31.90\% & 41.46\% & \textbf{49.23\%} & 87.40\% \\
 & TCN + Aug & 15.12\% & 25.77\% & 7.09\% & 39.74\% & 47.88\% & 18.20\% & 14.67\% & 6.04\% & 40.10\% \\
 & TCN + KL + Aug & \textbf{23.12\%} & 32.59\% & \textbf{77.27\%} & \textbf{73.94\%} & \textbf{94.87\%} & \textbf{34.23\%} & \textbf{42.79\%} & 47.78\% & \textbf{87.42\%} \\
 \hline
\multirow{4}{*}{MAE}
 & TCN & \textbf{29.34} & 41.49 & 16.59 & 6.93 & 29.94 & 40.06 & 47.22 & 17.32 & 54.65 \\
 & TCN + KL & 30.64 & \textbf{36.44} & 8.56 & 5.17 & 13.72 & 21.73 & \textbf{18.86} & 18.24 & \textbf{23.91} \\
 & TCN + Aug & 31.88 & 51.99 & 17.59 & 6.40 & 30.26 & 40.21 & 47.06 & \textbf{16.96} & 55.63 \\
 & TCN + KL + Aug & 35.16 & 40.63 & \textbf{8.49} & \textbf{2.84} & \textbf{12.61} & \textbf{20.63} & 20.65 & 18.46 & 23.95 \\
 \hline
\end{tabular}
\end{table*}

\begin{table}[!t]
\centering
\caption{Computational Complexity and Inference Efficiency}
\label{tab:complexity}
\begin{tabular}{@{}lccc@{}}
\toprule
\textbf{Model} & \textbf{Params} & \textbf{MFLOPs} & \textbf{CPU (ms)} \\ \midrule
NILMFormer & 383,283 & 1357.1 & 39.35 \\
BERT4NILM & 1,942,529 & 2166.8 & 26.60 \\
BiLSTM & 17,149,648 & 676.0 & 17.76 \\
BiGRU & 243,882 & 485.8 & 55.89 \\ \midrule
\textbf{Proposed} & \textbf{31,009} & \textbf{61.6} & \textbf{2.78} \\ \bottomrule
\end{tabular}
\end{table}

\begin{figure}[!t]
\centering
\includegraphics[width=0.9\columnwidth]{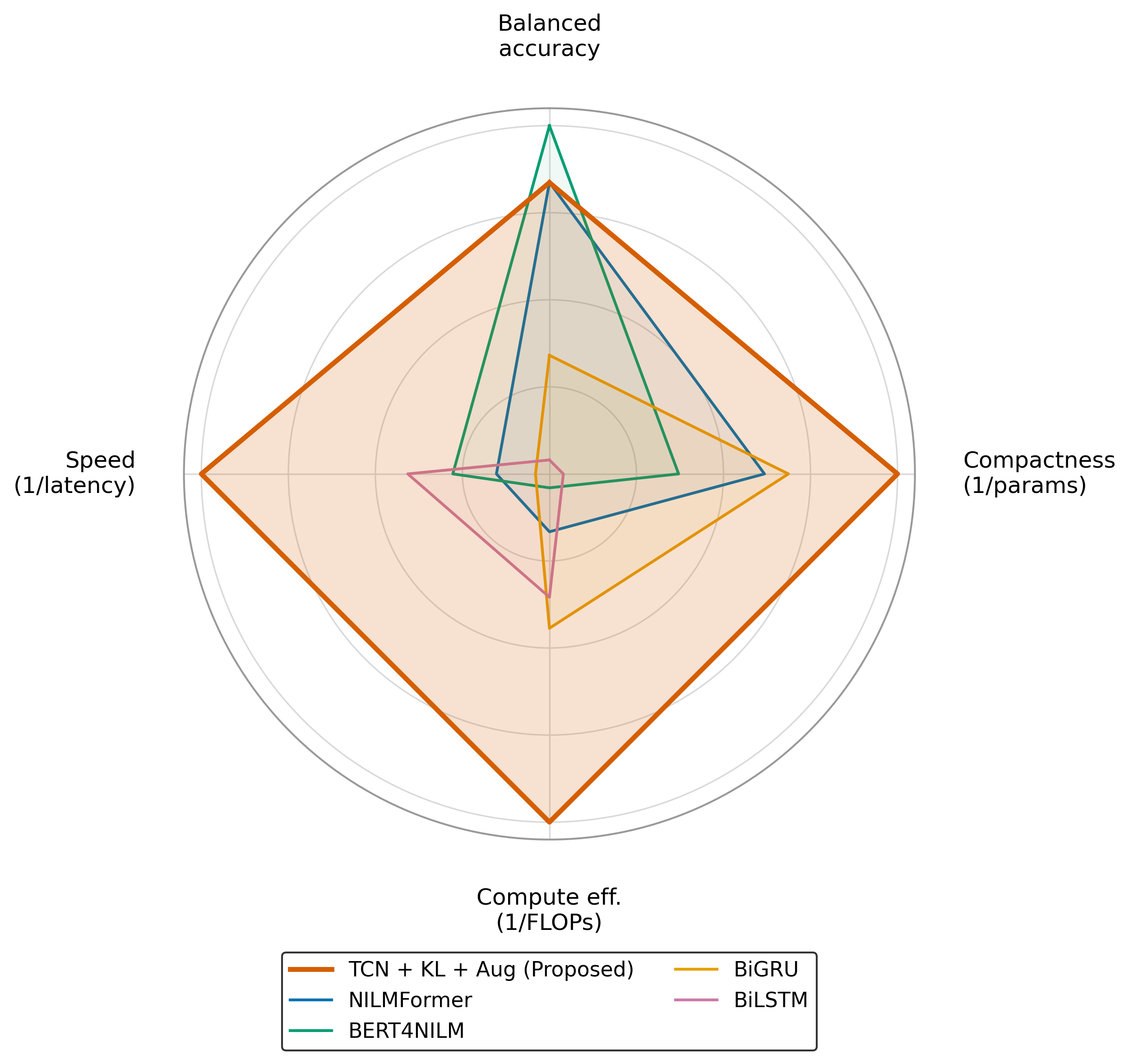}
\caption{Accuracy--efficiency trade-off.}
\label{fig:radar}
\end{figure}

The experiments were conducted on a workstation equipped with an NVIDIA RTX~6000 Ada GPU and an Intel Xeon w7-3565X CPU. The detailed hardware configuration and training hyperparameters are summarized in Table~\ref{tab:experimental_setup}. All models, proposed and baseline, are trained per appliance under the same pipeline, with identical splits, scaling, thresholds and loss, and with the same training schedule for every model within a dataset. The two datasets, however, use different stopping criterias. On UK-DALE, every model is trained for three epochs, so the results are a conservative lower bound for all models. On REDD, models are trained for up to 100 epochs with early stopping on the validation loss of a held-out validation house. Each configuration is trained with three window lengths (128, 256 and 512 samples) and three random seeds, and every value reported in Tables~\ref{table:performance_comparison} and~\ref{table:ablation_study} is the mean of these nine runs.

\subsection{Evaluation Metrics}

To comprehensively assess the disaggregation performance, the model is evaluated using both energy estimation metrics and state detection metrics, conforming to NILM benchmark standards.

\subsubsection{Power Estimation Accuracy}
The accuracy of the continuous power predictions is quantified using the Mean Absolute Error (MAE)

MAE measures the average absolute deviation between the predicted power $\hat{y}_k[n]$ and the ground truth power $y_k[n]$ for an appliance $k$ over a sequence of length $T$. It provides a direct interpretation of the average error in Watts:
\begin{equation}
    \text{MAE}_k = \frac{1}{T} \sum_{n=1}^{T} \left| \hat{y}_k[n] - y_k[n] \right|
\end{equation}

\subsubsection{State Detection Performance}

The state classification performance was evaluated using the balanced accuracy (BA) for the $k$-th appliance. Using the true positives (TP), true negatives (TN), false positives (FP) and false negatives (FN) of these temporal states, BA is defined for the $k$-th appliance as
\begin{equation}
    \text{BA}_k = \frac{1}{2}\left(\frac{\text{TP}_k}{\text{TP}_k+\text{FN}_k} + \frac{\text{TN}_k}{\text{TN}_k+\text{FP}_k}\right)
\end{equation}

Furthermore, appliances operate in an inactive (OFF) state for the majority of the temporal horizon, raw power metrics can obscure poor activation detection. We utilize the $F_1$-score to evaluate the model's ability to accurately detect ON/OFF states.

An appliance $k$ is considered active if its true or predicted power exceeds a predefined operational threshold $\theta_k$. The $F_1$-score calculates the harmonic mean of precision and recall:
\begin{equation}
    F_{1,k} = \frac{2 \cdot \text{TP}_k}{2 \cdot \text{TP}_k + \text{FP}_k + \text{FN}_k}
\end{equation}
The $F_1$-score effectively penalizes both missed activations and false-positive ON detections, providing a robust measure of temporal alignment in highly skewed load profiles.

\subsection{Datasets and Baselines}
To comprehensively evaluate the disaggregation performance, we utilized two widely adopted standard benchmark datasets: UK-DALE and REDD. The datasets were partitioned into training and testing sets using a leave-one-house-out (LOHO) strategy to ensure that the models are evaluated on unseen household power topologies. The exact data splits are detailed in Table \ref{tab:dataset_splits}.

For each dataset, the appliance activations used for synthetic data generation, the KLT basis and the scaling constant $C$ were all derived from the training houses only; the held-out test house was used exclusively for evaluation. In both training and testing, the aggregate is the sum of the target appliances' submeter readings (five for UK-DALE, four for REDD) rather than the measured mains.

All experiments were conducted on the appliances listed in Table \ref{tab:thresholds}. The threshold values used to determine each appliance's active (ON) state were adopted from \cite{10.1145/2821650.2821672}.

\subsection{Experimental Results}

\subsubsection{Overall Disaggregation Performance}
Table \ref{table:performance_comparison} presents the BA, F1-score, and MAE across the target appliances for both the UK-DALE and REDD datasets, each value being the mean of nine runs. No single method is best on every appliance and metric; the proposed method is competitive with the sequence-based (BiLSTM, BiGRU) and attention-based (NILMFormer, BERT4NILM) baselines while being far smaller and faster (Section~\ref{subsec:complexity}).

For the UK-DALE dataset, the proposed framework achieved the lowest MAE on the Microwave (2.84 W), outperforming the next best baseline (BERT4NILM at 3.83 W) and traditional sequence models (BiLSTM at 6.05 W). In terms of state classification, the proposed model achieved strong performance on the Fridge, with a BA of 95.69\% and an F1 of 94.87\%, indicating a reasonable capacity to identify both the active duration and power draw of low-frequency periodic loads.

On REDD, the proposed architecture yielded competitive or the highest balanced accuracies, demonstrating resilience against data scarcity and unseen household environments. For instance, in Microwave disaggregation on REDD, the proposed method achieved a BA of 92.51\%, against 85.39\% for NILMFormer and 84.23\% for BERT4NILM.

\subsubsection{Appliance-Specific Analysis}
To understand the architectural advantages, we illustrated model behavior by comparing ground-truth power signatures against predicted outputs for a refrigerator in UK-DALE and REDD (Fig.~\ref{fig:sota-comparison}). On UK-DALE, the proposed method accurately traces the switch-on and switch-off of each cycle, although it underestimates their level, whereas BiGRU misses the cycles and BiLSTM produces scattered spikes. REDD is more difficult for every model: the proposed method detects both cycles but underestimates them, BERT4NILM recovers the first ON period at roughly half its true power, and BiGRU largely misses these cycles while BiLSTM produces noisy estimates.

\subsubsection{Ablation Study}
\label{subsec:ablation_study}

To identify the contribution of each proposed component of the pipeline, we evaluated an ablation on the KLT front end and synthetic augmentation stages. The results are presented in Table \ref{table:ablation_study}.

The KLT front end accounts for nearly all of the gain. Averaged over appliances, adding it to the vanilla TCN raises balanced accuracy by 18.8 points and F1 by 32.0 points and lowers MAE by 6.0~W on UK-DALE; on REDD, it raises balanced accuracy by 22.2 points and F1 by 30.6 points and lowers MAE by 19.1~W. However, replacing the real training data with synthetic aggregates has a much smaller effect.

On UK-DALE the full model is the strongest variant in most columns, while on REDD it is closely matched by the KLT-only variant (six versus four of the twelve columns). The role of augmentation is therefore less to raise accuracy than to allow the model to be trained without real aggregate recordings, which enables the update pathway of Fig.~\ref{fig:proposed_model_update_pipeline}.

\subsubsection{Computational Complexity}
\label{subsec:complexity}
Table \ref{tab:complexity} compares the computational complexity and inference speed of the models evaluated. The proposed model contains 31,009 parameters, fewer than all four baselines. It also requires only 2.78 ms per 1000 samples on a single CPU thread, 84--95\% lower latency than the four baselines evaluated. These results support the framework's lightweight, edge-oriented design, though as workstation-measured figures they remain comparative, and embedded validation is left to future work.

Fig.~\ref{fig:radar} summarizes this trade-off: the proposed model is the most efficient on every computational axis while remaining close to the most accurate baselines.

%% file: src/4_Conclusion.tex
\section{Conclusion}

This work aims to resolve two critical bottlenecks in practical low-frequency smart grid deployment: data scarcity and computational constraints. Rather than relying on large, resource-intensive datasets, we proposed a boundary-aware synthetic data generation pipeline that cuts measured appliance traces only within inactive regions, and showed that the model can be trained entirely on these synthetic aggregates with only a small effect on accuracy. Furthermore, an empirical Karhunen-Lo\`eve Transform (KLT) filter bank supplies decorrelated, appliance-informed features to a compact temporal convolutional network (TCN), and accounts for most of the accuracy gain in our ablation. This reduces the need for heavily parameterized networks in low-frequency load monitoring.

Leave-one-house-out evaluations on UK-DALE and REDD at 10-s sampling show that the proposed architecture achieves comparable or superior accuracy at a much lower computational cost. The framework is competitive with recent sequence and attention-based models across energy estimation (MAE) and state detection (BA, F1) metrics. Each appliance model has only 31,009 parameters and needs 2.78~ms per 1000 samples on a single CPU thread, 84--95\% less than the evaluated baselines. Future research will focus on the physical implementation of the proposed KLT-TCN architecture onto resource-constrained smart meter hardware to benchmark real-time latency. Additionally, we aim to extend the automated synthetic augmentation engine to model highly non-stationary, multi-state appliances across different geographical regions.